\documentclass[notitlepage,prd,twocolumn,showpacs,preprintnumbers,nofootinbib,superscriptaddress]{revtex4-1}
\usepackage{amssymb,amsmath,placeins,epsfig,array}
\usepackage[usenames,dvipsnames]{color}
\usepackage[normalem]{ulem}
\usepackage[breaklinks,colorlinks,urlcolor=blue,citecolor=blue,linkcolor=blue]{hyperref}
\usepackage{lipsum}

\def\gsim{\mathrel{\raise.3ex\hbox{$>$\kern-.75em\lower1ex\hbox{$\sim$}}}}
\DeclareMathOperator{\ev}{eV}  \DeclareMathOperator{\mev}{MeV}               \DeclareMathOperator{\kel}{K}
         \newcommand{\cO}{{\cal O}}    
\newcommand{\ep}{\epsilon}  
\newcommand{\ie}{{\it i.e.~}}  \newcommand{\eg}{{\it e.g.~}}
   \def\oL{\overline} 
\newcommand{\pL}{\left(} \newcommand{\pR}{\right)} \newcommand{\bL}{\left[} \newcommand{\bR}{\right]}    
\newcommand{\beq}{\begin{equation}} \newcommand{\eeq}{\end{equation}}
\newcommand{\bea}{\begin{eqnarray}} \newcommand{\eea}{\end{eqnarray}}

\newcommand{\Eq}[1]{Eq.~(\ref{#1})} \newcommand{\Eqs}[2]{Eqs.~(\ref{#1}) and (\ref{#2})} 
  
\newcommand{\Fig}[1]{Fig.~\ref{#1}} 

\newcommand{\App}[1]{App.~\ref{#1}}
\newcommand{\Neff}{N_{\rm eff}}
\newcommand{\DNeff}{\Delta N_{\rm eff}}

\begin{document}

\preprint{FERMILAB-PUB-19-565-T}

\title{The Cosmological Evolution of Light Dark Photon Dark Matter}
\author{Samuel D.~McDermott}
\affiliation{Theoretical Astrophysics Group, Fermi National Accelerator Laboratory, Batavia, IL, USA}
\author{Samuel J.~Witte}
\affiliation{Instituto de F\'{\i}sica Corpuscular (IFIC), CSIC-Universitat de Val\`encia, Spain}

\begin{abstract}
Light dark photons are subject to various plasma effects, such as Debye screening and resonant oscillations, which can lead to a more complex cosmological evolution than is experienced by conventional cold dark matter candidates. Maintaining a consistent history of dark photon dark matter requires ensuring that the super-thermal abundance present in the early Universe {\emph{(i)}} does not deviate significantly after the formation of the CMB, and {\emph{(ii)}} does not excessively leak into the Standard Model plasma after BBN. We point out that the role of non-resonant absorption, which has previously been neglected in cosmological studies of this dark matter candidate, produces strong constraints on dark photon dark matter with mass as low as $10^{-22}\ev$.  Furthermore, we show that resonant conversion of dark photons after recombination can produce excessive heating of the IGM which is capable of prematurely reionizing hydrogen and helium, leaving a distinct imprint on both the Ly$-\alpha$ forest and the integrated optical depth of the CMB. Our constraints surpass existing cosmological bounds by more than five orders of magnitude across a wide range of dark photon masses.
\end{abstract}
\maketitle

\section{Introduction}
As many once-favored  models of particle dark matter become increasingly constrained (see \eg~\cite{Escudero:2016gzx,Arcadi:2017kky,Balazs:2017ple,Roszkowski:2017nbc,Blanco:2019hah}), candidates other than those resulting from weak-scale thermal freeze-out have been the subject of growing focus and development. One candidate of recent interest is the dark photon, $A'$ \cite{Jaeckel:2008fi,Pospelov:2008jk,Redondo:2008ec,Mirizzi:2009iz,Nelson:2011sf,Arias:2012az,Graham:2015rva,Agrawal:2018vin, Dror:2018pdh, Co:2018lka, Bastero-Gil:2018uel, Long:2019lwl,AlonsoAlvarez:2019cgw,Nakayama:2019rhg}, which arises from an abelian group outside of the Standard Model (SM) gauge group. This particle may ``kinetically mix'' with the SM photon via the renormalizable operator $\ep \, F^{\mu \nu} \, F'_{\mu \nu}\, / \, 2$ \cite{Holdom:1985ag}, with `natural' values of $\epsilon$ typically ranging from $10^{-16}$ to $10^{-2}$~\cite{Dienes:1996zr,Abel:2003ue,Abel:2006qt}.

Historically, one of the more problematic features of light vector dark matter has been the identification of a simple, well-motivated production mechanism. Early work on the subject suggested that such a candidate could be produced via the misalignment mechanism~\cite{Nelson:2011sf}, similar to that of axion dark matter (see \eg~\cite{Marsh:2015xka,Irastorza:2018dyq}), but it was later pointed out that this mechanism is inefficient at generating the desired relic abundance unless one also introduces a large non-minimal coupling to the curvature $\mathcal{R}$~\cite{Arias:2012az,Graham:2015rva,AlonsoAlvarez:2019cgw}. Such a coupling, however, can introduce ghost instabilities in the longitudinal modes~\cite{Himmetoglu:2008zp,Himmetoglu:2009qi,Karciauskas:2010as}; while it may be possible to avoid this feature, proposed solutions come at the cost of additional model complexity~\cite{Nakayama:2019rhg}. The work of \cite{Graham:2015rva} provided a compelling alternative production mechanism due to fluctuations of the metric during a period of early-universe inflation, but the non-observation of primordial gravitational waves constrain this mechanism from producing a viable dark matter population if $m_{A'} \lesssim \mu$eV. More recently, \cite{Agrawal:2018vin, Dror:2018pdh, Co:2018lka, Bastero-Gil:2018uel, Long:2019lwl} showed that a dark photon coupled to a hidden sector (pseudo)scalar field can generate the entire dark matter with masses as light as $m_{A'} \sim 10^{-20} \ev$. This super-thermal population of dark photons is generated by temperature-dependent instabilities or defects in the (pseudo)scalar field.  Given that various works have now provided more compelling mechanisms to generate what had perhaps previously been a more speculative dark matter candidate, we find it timely to revisit old, and develop novel, cosmological constraints on (and potential signatures of) light dark photon dark matter. 

The observational signatures of dark photon dark matter are quite distinct from canonical weak-scale particles. Various cosmological effects of light dark photon dark matter have been investigated over the years, typically focusing exclusively on the observational consequences arising from the resonant transition between dark and visible photons that occurs when the plasma frequency $\omega_p$ is approximately equal to the mass of the dark photon $m_{A'}$ \cite{Arias:2012az}. These constraints, however, are typically only applicable for $m_{A'} \geq \bar{\omega}_p^{0} \sim 10^{-14}$ eV, $\bar{\omega}_p^{0}$ being the background plasma frequency today. More recently, limits on very light dark photons were obtained using the observation that the kinetic mixing allows for an off-shell (non-resonant) absorption of dark photons, subsequently heating baryonic matter; if this heating is sufficiently large, it may destroy the thermal equilibrium of the  Milky Way's interstellar medium~\cite{Dubovsky:2015cca}, that of ultra-faint dwarf galaxies such as Leo T~\cite{Wadekar:2019xnf}, or cold gas clouds in the Galactic Center~\cite{Bhoonah:2018gjb}. This idea has also been used to project the sensitivity that could be obtained from future 21 cm experiments which observe absorption spectra during the cosmic dark ages~\cite{Kovetz:2018zes}.

In this work, we put forth a simple cosmological picture of dark photon dark matter, requiring only that $\emph{(i)}$ dark matter is not overly depleted after recombination and $\emph{(ii)}$ the energy deposited into the SM plasma does not produce unwanted signatures in BBN, the CMB, or the Ly-$\alpha$ forest. We identify (and describe in a unified manner) the resonant and non-resonant contributions to both of these classes of observables. We find that these simple and robust requirements lead to extremely stringent constraints for light photon dark matter, covering dark photon masses all the way down to $\sim 10^{-22}$ eV. Our constraints are stronger than existing bounds across a wide range of masses (in some cases by more than five orders of magnitude), and are robust against astrophysical uncertainties\footnote{We choose here to neglect bounds from superradiance which in principle could constrain dark photons with masses below $\sim 10^{-11}$ eV~\cite{Baryakhtar:2017ngi}, as the existence of such bounds require self-interactions of the new gauge boson to be small~\cite{Agrawal:2018vin}. }.

\begin{figure*}[t]
\begin{center}
\includegraphics[width=0.85\textwidth]{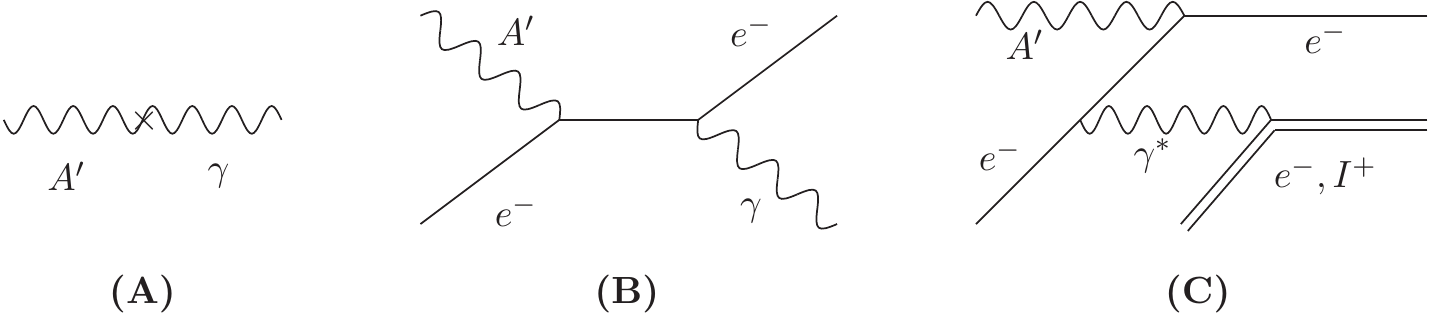}
\caption{Processes by which photons and dark photons interconvert. Since the dark matter is inherently cold, the processes labelled {\bf (A)} and {\bf (B)} require $m_{A'} \geq \omega_p$. In the case of inverse bremsstrahlung, shown in panel {\bf (C)}, the fact that the photon can be off-shell allows dark matter to be absorbed even when  $m_{A'} \ll \omega_p$. We include these diagrams to provide the reader with intuition, but use the formalism described in text for all computations. }
\label{feyn-diag}
\end{center}
\end{figure*}

This work is organized as follows. We begin by outlining the relevant on- and off-shell conversion processes that alter the energy and number densities of the dark sector and SM plasma. We then discuss various cosmological implications for the existence of light dark photon dark matter, including modifications to the evolution of the energy density after neutrino decoupling, spectral distortions produced in the CMB, dark matter evaporation, and modifications to the Ly-$\alpha$ forest from the heating of the IGM. We conclude by discussing more speculative ways in which sensitivity can be extended to the low mass regime.

\section{Plasma Mass and (Dark) Photon Conversion}
Dark photons and SM photons can interconvert through cosmic time. Accurately treating this conversion requires accounting for plasma effects: the SM photon has a modified dispersion relation in a charged plasma, given by $\omega^2 = {\rm Re}\Pi(\omega, k, n_e) + k^2$. The dimensionful scale that governs the SM photon dispersion relation is the plasma mass ${\rm Re}\Pi(\omega, k, n_e) \propto \omega_p^2(z) = 4\pi\alpha_{EM}\sum n_i(z) / E_{F,i}$; here, $n_i$ is the number density of species $i$ and $E_{F,i} = \sqrt{m_i^2 + (2\pi^2 n_i)^{2/3}}$ is the charged particle Fermi energy. We will focus on cosmological epochs for which the only relevant species is the electron, with number density given by 
\begin{equation}\label{electron-density}
n_e = X_e(z) \left(1 - \frac{Y_p}{2}\right) \eta \frac{2 \zeta(3)}{\pi^2} T_0^3 (1+z)^3 \, .
\end{equation}
In \Eq{electron-density}, $X_e(z)$ is the free electron fraction, $Y_p$ is the primordial helium abundance, $\eta$ is the baryon to photon ratio, and $T_0$ is the temperature of the CMB today. The function $X_e(z)$ can be obtained using the open source code {\tt class}~\cite{Blas:2011rf}, and we fix $Y_p= 0.245$~\cite{Aver:2015iza,Pitrou:2018cgg} and $T_0=2.7255$ K~\cite{Fixsen:2009ug}.

In general, dark photons and SM photons will convert with equal probability. An asymmetry in energy flow is therefore possible only due to initial conditions: at the time of the formation of the CMB the SM photons are described to good precision by a blackbody at a temperature $T_0(1+z_{\rm CMB})$, while dark photons that constitute the cold dark matter must be a collection of non-thermal particles with a number density far larger than $n_\gamma$ and an energy spectrum peaked very close to $m_{A'}$ (for the sake of completeness, we will also address the possible existence of dark photons with a very small initial number density). The total energy taken from the reservoir of cold dark photons and introduced to the SM photon bath is 
\beq \label{A'-to-gamma}
\Delta \rho_{A' \to \gamma} =  \int dz \, P_{A' \to \gamma}(z)  \times \rho_{A'}(z) \, ,
\eeq
where $P_{A' \to \gamma}(z)$ is the redshift-dependent probability of conversion from an $A'$ to a SM photon and $\rho_{A'}(z)$ is the redshift dependent energy density of dark photons. Later, we will consider the energy injected normalized to the number density of baryons, which is given by \Eq{A'-to-gamma} with the simplifying substitution $\rho_{A'}(z) \to \rho_{A'}(z)/n_b(z)$. If the conversion probability is small, one can approximate $\rho_{A'}(z) \sim (1+z)^3 \rho_{A'}^0$, with $\rho_{A'}^0$ being the mean dark matter density today; however, in some cases, the probability is sufficiently large that dark matter density prior to conversion is significantly greater than the dark matter density after, in which case the aforementioned approximation is not valid.

Similarly to \Eq{A'-to-gamma}, we may write the energy extracted from the SM photon bath as~\cite{Jaeckel:2008fi,Mirizzi:2009iz}
\beq \label{gamma-to-A'}
\Delta \rho_{\gamma \to A'}(E) = \frac{T_0^4}{\pi^2}  \int dzdx \, \frac{x^3 (1+z)^4}{e^{x} - 1} P_{\gamma \to A'}(x,z) ,
\eeq
where $x \equiv E / T$, and we have explicitly included the energy dependence in the conversion probability since the CMB spectrum is far broader than that of cold dark matter, and is well-measured near the peak. 

We will use \Eqs{A'-to-gamma}{gamma-to-A'} to constrain the existence of dark photons. As we show below, the most sensitive probes are from limits on the heating of the SM bath after recombination. Before deriving these bounds, we first discuss the different routes by which a dark photon can convert to a SM photon.

\section{On-Shell and Off-Shell Conversion}
A dark photon can convert either to an on-shell SM photon (via oscillation or 2-to-2 processes) or to a virtual SM photon (through a 3-to-2 process). Examples are shown in \Fig{feyn-diag}. While the 3-to-2 process is na\"ively negligible due to the extra phase space and the factor of $\alpha_{\rm EM}$, it can dominate in some regimes of parameter space, depending on kinematic matching considerations.

The on-shell processes of interest are oscillation and semi-Compton absorption. These can operate efficiently if $m_{A'} \gtrsim \omega_p$, but $A' \to \gamma$ is strongly suppressed for a cold dark photon bath if $m_{A'} < \omega_p$. On-shell phenomena are most pronounced at a level crossing, occurring at $m_{A'} \simeq \omega_p(z)$ for traverse modes and $\omega \simeq \omega_p(z)$ for longitudinal modes. In practice, these occur at the same redshift for on-shell conversion of dark photon dark matter, since $\omega \simeq m_{A'}$; note that this need not be true for off-shell conversion or for conversion to non-cold dark photons. The probability of a transition at the time of level crossing is governed by the non-adiabaticity of the change in $\omega_p(z)$, and is approximately given by the Landau-Zener expression \cite{Parke:1986jy, Kuo:1989qe, Mirizzi:2009iz}
\begin{equation} \label{LZp}
P_{A' \to \gamma}^{\rm (res)} \simeq \frac{\pi \, \epsilon^2 \, m_{A'}^2 }{\omega \, (1+z) \, H(z)}\left| \frac{d\log{\omega_p^2(z)}}{dz}\right|^{-1}\,  \delta(z-z_{\rm res}).
\end{equation}
\Eq{LZp} is valid only when $P_{A' \to \gamma} \ll 1$. When this condition is violated we adopt the general expression, which can be found \eg~in~\cite{Mirizzi:2009iz,Arias:2012az}. The delta function in \Eq{LZp} makes the redshift integral in \Eq{A'-to-gamma} trivial. A similar expression holds for resonant $\gamma \to A'$ conversion.

In contrast to resonant conversion, an off-shell process like inverse bremsstrahlung will operate even for $m_{A'} \ll \omega_p$, and can dominate the heating rate despite entering at a lower order in $\alpha_{\rm EM}$. This process can occur off resonance and is not forbidden by energy conservation because the outgoing photon is not on-shell. This process leads to a heating of the plasma proportional to the number of dark matter particles absorbed. As described in \cite{Dubovsky:2015cca}, this process is subject to Debye screening when $m_{A'} \neq \omega_p$, and thus the rate of loss of energy from the cold dark photon reservoir is given by
\begin{equation} \label{Q-dp}
P_{A' \to \gamma}^{\rm (nonres)}  \simeq  \frac{ \ep^2 \nu}{2(1+z) \, H(z)} \bL \frac{m_{A'}^2}{\omega_p(z)^2} \bR^{{\rm sign}[\omega_p(z)-m_{A'}]} \, ,
\end{equation}
with the frequency of electron-ion collisions $\nu$ given by
\begin{equation}
\nu = \frac{4\, \sqrt{2\pi} \, \alpha_{\rm EM}^2 \, n_e}{3\, \sqrt{m_e \, T_e^3}}\,  \log \pL \sqrt{ \frac{4 \pi \, T_e^3}{\alpha_{\rm EM}^3 \, n_e} }\pR \, .
\end{equation}
The fact that \Eq{Q-dp} is proportional to $\nu$ is related to the fact that this is an inherently off-shell process. This rate decouples like $(\ep m_{A'}/\omega_p)^2$ for $m_{A'}< \omega_p$ (and, conversely, like $(\ep \omega_p/m_{A'})^2$ for $m_{A'}> \omega_p$), but even an arbitrarily light dark photon may participate, and the rate does not abruptly drop to zero.

In the following, we derive constraints on the kinetic mixing parameter for light to ultra-light dark photons, assuming either that dark photons do or do not comprise the entirety of dark matter. We analyze both resonant and non-resonant processes that lead to either a deposition of energy into or removal of energy from the SM plasma. By including off-shell dark photon absorption, we find that there exist stringent cosmological bounds on the kinetic mixing of the dark photon dark matter at all relevant masses.

\section{Pre-CMB Considerations}
\begin{figure}
\includegraphics[width=0.5\textwidth]{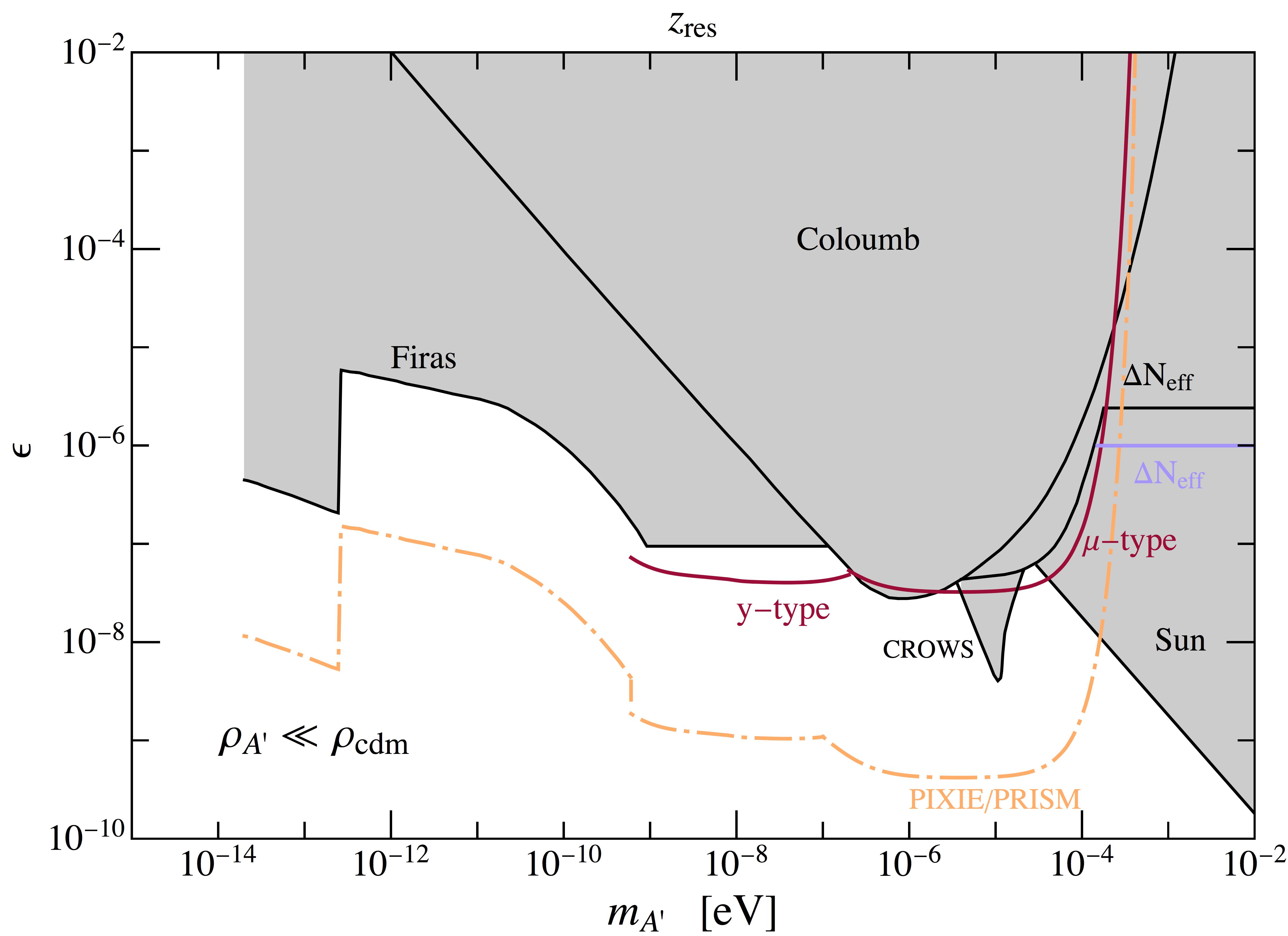}
\caption{Bounds that apply for low (or zero) initial abundance of dark photons arising from constraints on $\mu$- and $y$-type distortions using the Green's function formalism of~\cite{Chluba:2013vsa,Chluba:2015hma}. Also shown are existing constraints from spectral distortions~\cite{Mirizzi:2009iz}, $5^{\rm th}$ force experiments~\cite{Williams:1971ms,Bartlett:1988yy}, modifications to $\DNeff$~\cite{Jaeckel:2008fi}, stellar cooling constraints~\cite{An:2013yfc,Redondo:2013lna,Vinyoles:2015aba}, and the CROWS experiment~\cite{Betz:2013dza}. Finally, we project the sensitivity of experiments like PIXIE and PRISM to $\mu$- and $y$-type distortions (similar bounds have been found in~\cite{Kunze:2015noa}). The redshift for which a dark photon with mass $m_{A'}$ undergoes resonant conversion $z_{\rm res}$ is shown on the top x-axis for comparison (neglecting reionization).}
\label{fig:no_DM}
\end{figure}

Resonant conversions between photons and dark photons at temperatures $T \lesssim \mathcal{O}({\rm MeV })$ and prior to recombination can leave discernible signatures in the energy density inferred from BBN and the CMB. In the absence of a dark photon population, CMB photons will resonantly convert and populate a relativistic dark sector, producing a positive shift in the effective number of light degrees of freedom $\Neff$. Such a bound was first derived in~\cite{Jaeckel:2008fi}, and is reproduced in \Fig{fig:no_DM}.

Alternatively, should dark photons contribute significantly to the cold dark matter energy density, conversions from the dark sector into the SM photon bath will be the more efficient process (owing to the large dark photon number density, and the fact that low-energy photons with $\omega \ll T$ can be produced). In fact, resonant production of photons can be so efficient that nearly all of the dark matter can be converted into radiation. Na\"ively this appears problematic for the existence of dark matter today; however, the earliest measurement of cold dark matter energy density comes from the CMB, and the matter energy density before this time is basically unconstrained. For this scenario to remain consistent with observations, one may postulate the existence of an initial population of cold dark photons much larger than what would be expected given a $(1+z)^3$ extrapolation of $\Omega_{\rm CDM}^0$. Since the energy density of radiation redshifts more quickly than that of cold dark matter, one must also be concerned about the possibility of having a period of early matter domination during BBN. In order to ensure a successful nucleosynthesis, we require the initial matter density at $T \sim \mev$ to be no larger than the energy density stored in new effective light degrees of freedom, which are constrained during this epoch to be $\DNeff^{\rm(BBN)} \lesssim 0.5$~\cite{Berlin:2019pbq}. This constraint was first derived in~\cite{Mirizzi:2009iz}, and since it is logarithmically sensitive to the constrained value of $\DNeff$, the bounds derived here are effectively identical to those obtained nearly a decade ago.     

Remaining consistent with the thermal history as inferred from measurements of BBN and the CMB produces the strongest bounds on the kinetic mixing for values of the dark photon mass $m_{A'} \sim 10^{-4}$ eV. We derive the bounds shown in \Fig{fig:no_DM} and \Fig{fig:DM} using the latest constraints on $\DNeff$ from {\tt Planck}~\cite{Aghanim:2018eyx} and BBN~\cite{Cyburt:2015mya,Pitrou:2018cgg}.

\section{CMB Spectral Distortions}
Light dark photons depositing energy in the SM plasma at $z \lesssim 2\times10^{6}$ (\ie temperatures $T \lesssim 500$ eV) will produce distortions in the CMB blackbody spectrum. For redshifts $z \gtrsim 2\times 10^6$, double Compton (DC) scattering and bremsstrahlung are efficient at producing low energy photons which are subsequently up-scattered via Comptonization (see \eg~\cite{Chluba:2011hw,Khatri:2012tw,Chluba:2013vsa,Tashiro:2014pga} for an overview). This process of thermalization erases any spectral distortions that could arise as a result of the energy injection from dark sectors, and, because thermal equilibrium dictates the number density of photons as well as their spectrum, spectral distortions are possible only after photon-number-changing processes become inefficient. We provide a review of the signatures imprinted on the CMB from energy transfers between the dark and visible sectors in the Appendix, and focus below only the formalism adopted for computing the current limits and projected sensitivity.

Spectral distortions are constrained by various experiments, most notably COBE/FIRAS~\cite{Fixsen:1996nj}, to the level of $|y| \leq 1.5 \times 10^{-5}$ and $|\mu|\leq 6 \times 10^{-5}$~\cite{Tashiro:2014pga}. Future experiments such as PIXIE~\cite{Kogut:2011xw} and PRISM~\cite{Andre:2013afa,Andre:2013nfa} could enhance the sensitivity of these spectral distortions to the level of $|y|, \,|\mu| \lesssim 10^{-8}$.  Should dark photons not contribute to the dark matter, blackbody photons can resonantly convert and lead to a depression of the spectrum at the measured frequencies~\cite{Mirizzi:2009iz}. The analysis performed in~\cite{Mirizzi:2009iz}, however, focuses only on resonant conversions occurring in the frequency band observable by FIRAS. The bound derived using this method is clearly conservative, as conversions at frequencies below what is observable by FIRAS still occur, and for $z \gtrsim 10^3$  can still induce spectral distortions since Compton and bremsstrahlung processes are still partially active and lead to a modification of the blackbody spectrum. Similarly, should dark photons account for the entirety of dark matter, the energy deposited in the SM plasma will create $\mu$- and/or $y$-type distortions, depending on when this process takes place  (see Appendix to understand for which redshifts energy deposition results in $\mu$ and $y$-type distortions, and the effects they induce on the black body spectrum). Existing constraints were derived on this energy deposition in a heuristic way in~\cite{Arias:2012az}; here, we attempt provide a more detailed a rigorous analysis of this effect.

\begin{figure*}[t]
\begin{center}
\includegraphics[width=0.8\textwidth]{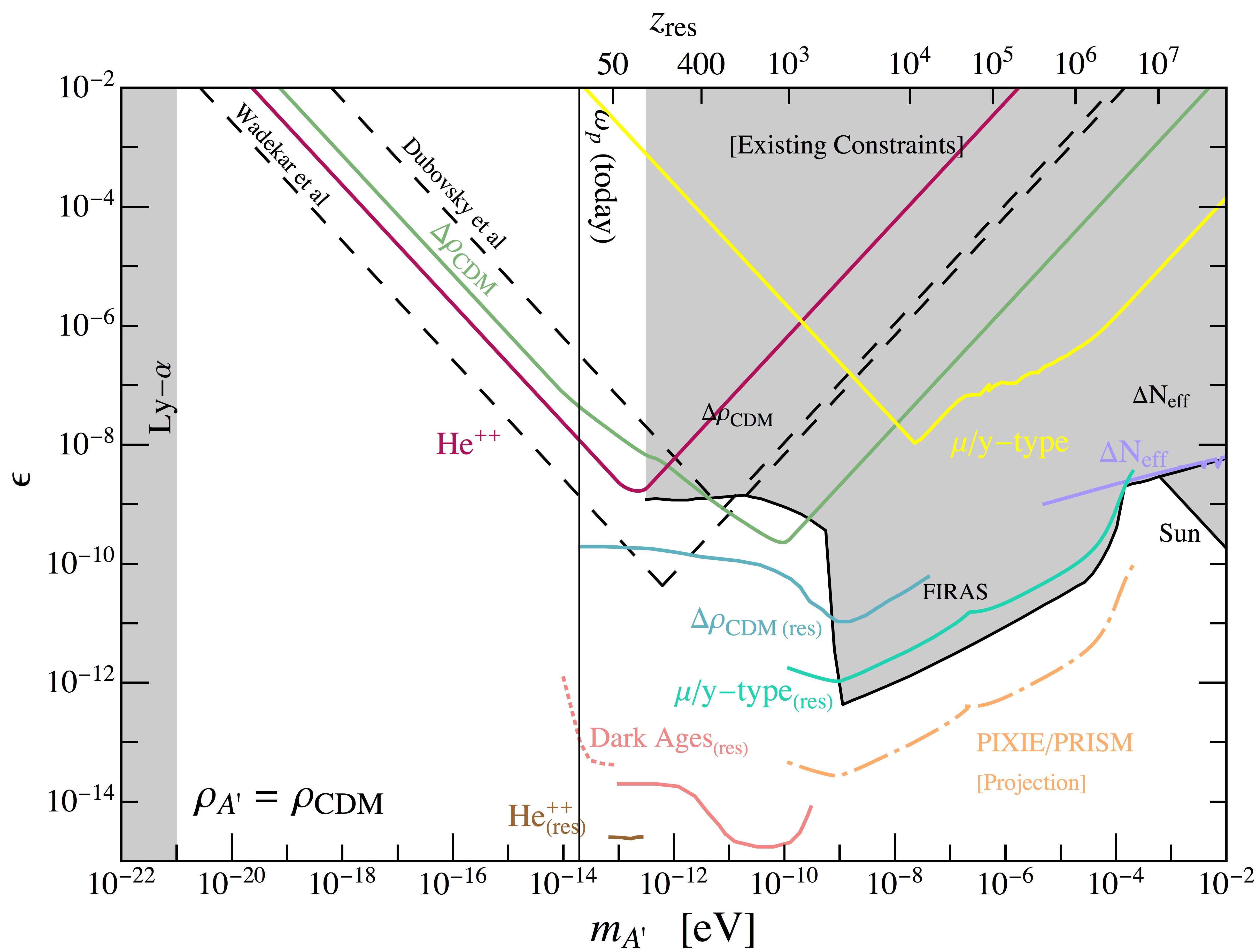}
\caption{Limits on dark photon dark matter from: $\Neff$ (purple);  $\mu$- and $y$-type distortions (resonant and non-resonant correspond to teal and yellow, respectively); the  depletion of dark matter at the level of $10\%$ (resonant and non-resonant correspond to blue and green, respectively), as in \Eq{dm-survival}; energy deposition during the cosmic dark ages (pink solid) and enhancements in the integrated optical depth produced by resonant conversions (pink dotted), as in \Eq{eq:tau}; and heating of the IGM around the epoch of helium reionization (resonant and non-resonant correspond to brown and red, respectively), as in \Eq{HeII}. Existing cosmological constraints on modifications to $\DNeff$ during BBN and recombination~\cite{Arias:2012az}, spectral distortions~\cite{Arias:2012az}, the depletion of dark matter~\cite{Arias:2012az}, stellar cooling~\cite{An:2013yfc,Redondo:2013lna,Vinyoles:2015aba}, and the Ly-$\alpha$ forest~\cite{Irsic:2017yje}, are shown in grey for comparison. Dashed black lines denote astrophysical bounds derived from thermodynamic equilibrium of gravitationally collapsed objects: the Milky Way~\cite{Dubovsky:2015cca} (labeled `Dubovsky et al') and the ultra-faint dwarf galaxy Leo T~\cite{Wadekar:2019xnf} (labeled `Wadekar et al')\footnote{A similar bound has been estimated using thermodynamic equilibrium of gas clouds in the Galactic Center~\cite{Bhoonah:2018gjb}}. The mean plasma frequency today is shown for reference with a vertical line, along with the redshift dependence of the plasma frequency, neglecting reionization, on the upper axis. We include alongside this publication an ancillary file outlining the strongest constraint for each dark photon mass in order to ease reproduction of our bounds.  }
\label{fig:DM}
\end{center}
\end{figure*}

We compute constraints on dark photons from both resonant and non-resonant energy deposition and extraction using the Green's function formalism~\cite{Chluba:2013vsa,Chluba:2013pya,Chluba:2015hma,Chluba:2016bvg}; the results of these analyses are summarized in  \Fig{fig:no_DM} and \Fig{fig:DM} for the case in which the initial dark photon density is $\sim 0$ or equal to that of dark matter, respectively. Existing constraints on $\mu$- and $y$-type distortions come from COBE/FIRAS, and we also project future bounds for a PIXIE/PRISM-like experiment.
Specifically, the level of spectral distortions can be accurately approximated by convolving the energy deposition rate with a series of visibility functions accounting for the fraction of injected energy that produces a particular type of distortion. These expressions are given by:
\begin{align}\label{eq:yeq}
y & \simeq  \frac{1}{4} \int \, \frac{\mathcal{J}_y(t)}{\rho_\gamma(t)} \, \frac{d \rho(t)}{dt} \, dt \\  \label{eq:mueq}
\mu & \simeq  1.401 \int \, \frac{\mathcal{J}_{bb}(t) \, \mathcal{J}_\mu(t)}{\rho_\gamma(t)} \, \frac{d \rho(t)}{dt} \, dt  \, ,
\end{align}
with $d\rho/dt$ the energy density injected to or extracted from the plasma per unit time (assumed to be given by either a delta function or by $d\rho/dz \times dz/dt = \rho_{cdm}(z) \times P_{A' \to \gamma}^{\rm (nonres)} \times dz/dt$, for the case of resonant and non-resonant conversion respectively), and the visibility functions $\mathcal{J}_i$ are given by
\begin{align}
\mathcal{J}_{bb}(t) & = {\rm Exp}\left[{-\left(\frac{z}{z_\mu}\right)^{5/2}}\right] \\
\mathcal{J}_y(t) & = \left[1 + \left(\frac{1+z}{6\times 10^4} \right)^{2.58} \right]^{-1} \\
\mathcal{J}_\mu(t) & = 1 - \mathcal{J}_y \, .
\end{align}
Here, $z_\mu = 1.98 \times 10^6 \, (\Omega_b h^2/0.022)^{-2/5} \left[(1-Y_p/2)/0.88 \right]^{-2/5}$ is the redshift at which DC begins to become inefficient. These equations are only valid for $z \gtrsim 10^3$, explaining the somewhat unphysical truncation of bounds derived from resonant transitions shown in \Fig{fig:no_DM} and \Fig{fig:DM} at $m_{A'} \simeq 10^{-9}\ev$. We confirm the existing bounds from the FIRAS instrument in the range $10^{-14}\ev \lesssim m_{A'} \lesssim 10^{-9}\ev$ \cite{Arias:2012az}, and we scale these to future sensitivity expected by PIXIE/PRISM. In the scenario that dark photons constitute the entirety of dark matter, we show for completeness in \Fig{fig:DM} constraints derived from non-resonant dark photon absorption, obtained by combining \Eq{Q-dp} with \Eqs{eq:yeq}{eq:mueq}.

\section{Dark Matter Survival}
After recombination, dark photon dark matter can be depleted via the processes shown in \Fig{feyn-diag}. The total change in the dark matter energy density is given by integrating \Eq{A'-to-gamma} using \Eqs{LZp}{Q-dp} from redshift 0 to $z\sim10^3$. Should this change in density be sufficiently high, the relative abundance of dark matter observed today would differ from the value inferred by observations of the CMB. Maintaining consistency with current observations requires, at a minimum, that the density of decaying dark matter particles changes by no more than $\simeq 2-3\%$ after matter-radiation equality \cite{Poulin:2016nat,Xiao:2019ccl}. We begin here by deriving a conservative bound, imposing that off-shell processes change the dark matter density by no more than $10\%$, \ie
\beq \label{dm-survival}
\Delta \rho_{A'}^{0 \leq z \leq 1000} \leq 0.1 \times \rho_{A'}^0 \, .
\eeq
A similar bound has been derived with on-shell (resonant) conversion for dark photon masses $m_{A'} \gtrsim 10^{-12.5}$ eV in~\cite{Arias:2012az}. At lower masses, the resonant bound can no longer be applied and the off-shell process becomes dominant, albeit with a increasing suppression due to Debye screening, exhibiting the expected decoupling behavior with respect to $m_{A'}$. In the case of the resonant conversion, we derive a more rigorous bound using the latest CMB observations by {\tt Planck}~\cite{Aghanim:2019ame}. Specifically, we modify {\tt class} to include an abrupt change in the dark matter energy density, modeled using a {\tt tanh} function of width $\Delta z = 1$\footnote{We demonstrate in the following section that the timescale over which dark photons undergo resonant absorption and are subsequently absorbed by the IGM is much less than $\Delta z = 1$, making this approximation conservative. }, and perform an MCMC using {\tt montepython}~\cite{Audren:2012wb}. Our combined likelihood includes the {\tt Planck-2018 TTTEEE+low$\ell$TT+lowE+lensing} likelihood~\cite{Aghanim:2019ame} and observations of baryonic acoustic oscillations (BAOs) from the 6DF galaxy survey~\cite{Beutler:2011hx}, the MGS galaxy sample of SDSS~\cite{Ross:2014qpa}, and the CMASS and LOWZ galaxy samples of BOSS DR12~\cite{Alam:2016hwk}. We have chosen to include the low redshift BAO likelihoods in this part of the analysis as this allows for a more robust determination of the energy density in cold dark matter at low redshifts, although it is important to note that the results obtained using exclusively the Planck likelihood are quite similar to those shown here. We adopt flat priors on $\log_{10} m_{A'}$ and $\log_{10} \epsilon$ in the range of $[-9, -14]$ and $[-12, -7]$, respectively. The resultant $2\sigma$ bound is significantly stronger than the off-shell constraint across all masses for which resonant conversions can occur.  Our bounds are much stronger than those of~\cite{Arias:2012az}, because the constraints there were obtained by requiring $\tau<$ 1, corresponding to a change of 65\% in the dark matter energy density after the formation of the CMB. However, this constraint is now known to be much too conservative; e.g., decaying dark matter must not be depleted by more than a few percent \cite{Poulin:2016nat,Xiao:2019ccl}, which is similar to the constraint we find from {\tt class} for resonant transitions of dark photons. Thus, our result for resonant conversion after the CMB epoch is stronger by a factor of $\sim \sqrt{65\%/1\%} \sim \cO(10)$.

Of particular interest in cosmology today is the so-called Hubble tension, which is a $4-6\, \sigma$ disagreement between the value of $H_0$ inferred using local measurements~\cite{Riess:2016jrr,Riess:2019cxk,Freedman:2019jwv,Bonvin:2016crt,Birrer:2018vtm,Wong:2019kwg} and that inferred from early Universe cosmology~\cite{Aghanim:2018eyx} (see also \eg \cite{Verde:2019ivm}). It has been pointed out that resolving this tension seems to require early Universe physics~\cite{Knox:2019rjx}, and in particular favors a modification to the energy density near the time of recombination. Given that this model is capable of generating an abrupt change in the matter density (and thus the expansion rate) at the time of recombination, it is natural to wonder whether the effect could address any outstanding discrepancies between early and late Universe cosmology. As we will show in the following section, the impact of the energy injection from this resonant conversion process actually produces constraints sufficiently strong so as to eliminate the possibility of an $\mathcal{O}(1)\%$ change in $\Omega_{\rm CDM}$, as would be necessary to noticeably impact the inferred value of $H_0$.  This can easily be seen in \Fig{fig:DM} by comparing the relative limits derived using $\Delta \rho$ and energy injection during the dark ages; constraints from the latter, being orders of magnitude stronger than those derived using exclusively $\Delta \rho$, clearly do not permit significant changes in the dark matter energy density, and thus cannot modify the evolution of the energy density around recombination as would be required to shift the inferred value of $H_0$.

\section{Energy Deposition During Cosmic Dark Ages}
The energy per baryon stored in the dark sector is, on average, greater than $10^9$ eV (\ie $\rho_{\rm CDM} / n_b \sim \Omega_{\rm CDM} / \Omega_{b} \times m_p \sim 5 \times 10^9$ eV). For most dark matter candidates, the relevant processes allowing energy flow into the SM sector decouple well before the formation of the CMB. In the case of the dark photon, however, resonant transitions can concentrate this energy in a narrow window, leading to enhanced observable effects. Specifically, if the energy is deposited in the SM plasma after recombination, the induced heating can raise the temperature of the gas above the threshold for the collisional ionization of hydrogen, and induce an early, albeit short-lived, period of reionization. This will affect the integrated optical depth of the CMB, currently measured by {\tt Planck} to be $\tau = 0.054 \pm 0.007$~\cite{Aghanim:2018eyx}. 

There are a number of potential concerns that must be addressed before introducing the relevant formalism for tracking the impact of heating, and subsequent ionization, produced from resonant dark photon conversions. First, it is important to address the fate of photons injected into the medium after recombination. We demonstrate below that the change in the optical depth with respect to redshift is large at the time of production, and thus it is valid to assume that these photons are absorbed instantaneously. Next, if the timescale of the resonance is large relative to the timescale for absorption and collisional ionization, there will be a back reaction that disrupts the resonant conversion. We will show that in fact this is never the case for the redshifts and parameter space of interest, and one can safely assume that the processes of resonant conversion, absorption, and collisional ionization, take place independently in this order.

Let us begin by discussing the fate of photons produced from the resonant conversion of non-relativistic dark photons. We are interested in studying the resonance that occurs in both the transverse and longitudinal modes when $\omega \simeq \omega_p \simeq m_{A'}$. Since both modes are on-resonance, both modes will be produced in appropriate ratios, \ie one-third longitudinal and two-thirds transverse. Longitudinal modes, however, don't propagate and are thus immediately absorbed by the plasma. For transverse modes, one must compute the optical depth along the direction of propagation in order to determine whether or not these photons can be treated with the on-the-spot approximation (\ie they are absorbed instantaneously). For the energies studied here ($E_\gamma \leq 10^{-2}$ eV), the relevant process dictating the mean free path of a resonantly produced photon is simply bremsstrahlung absorption (also known as free-free absorption). The integrated optical depth from production at $z_i$ to some final redshift $z_f$ is given by~\cite{Chluba:2015hma}
\begin{multline}
\tau_{BR}(E_\gamma, z) = \int_{z_f}^{z_i} \, dz \, \frac{\Lambda_{BR}(z, E_\gamma) (1 - e^{-E_\gamma / T_e(z)})}{(E_\gamma / T_e(z))^3} \\ \times \, \frac{\sigma_T \, n_e}{H(z) (1+z)} \,
\end{multline}
where $\Lambda_{BR} = (\alpha \lambda_c^3 / 2\pi \sqrt{6\pi}) \,n_p\, \theta_e^{-7/2} \, g_{BR}(E_\gamma)$ is related to the bremsstrahlung emissivity, and $T_e$ is the temperature of the plasma. Here, $\lambda_c$ is the electron's Compton wavelength, $\theta_e = T_e / m_e$, $n_p$ is the proton number density, and $g_{BR}$ is the bremsstrahlung Gaunt factor, which we take from \cite{draine2011physics} (see also~\cite{Weinberg:2019mai} for a more generalized treatment of soft bremsstrahlung processes). 

\begin{figure}[]
\includegraphics[width=0.5\textwidth]{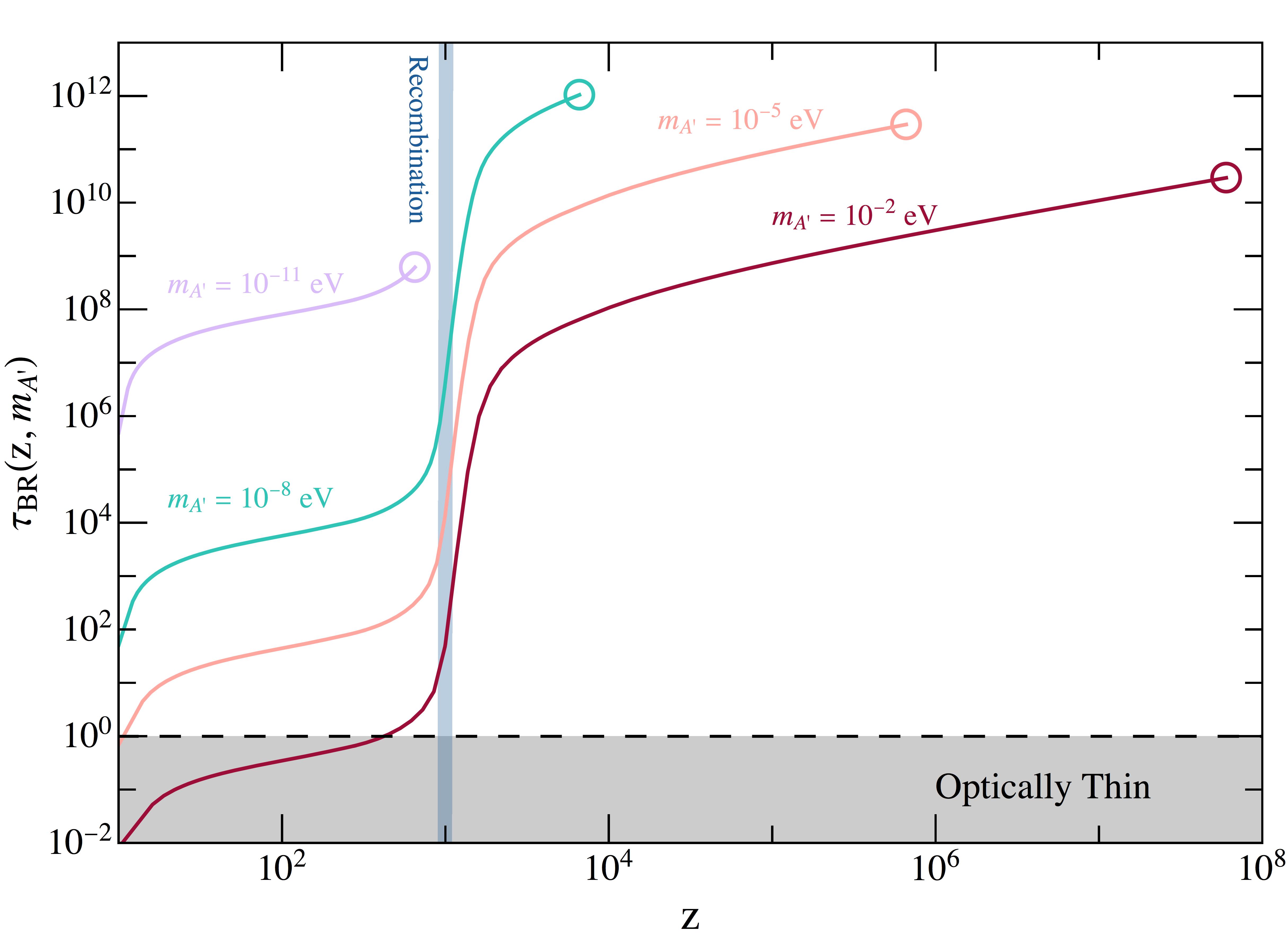}
\caption{Optical depth of visible photons resonantly produced from dark photons with masses $m_{A'}$, integrated from $z_f = 10$ to the redshift of production (denoted with colored circle). }
\label{fig:tau}
\end{figure}

In \Fig{fig:tau} we show the optical depth for a photon created with energy $m_{A'}$ at the redshift of resonance (\ie we take $\omega_p(z_i) = m_{A'}$) and taking $z_f = 10$. We adopt $z_f = 10$ rather than \eg~$z_f = 0$ because the post-reionization epoch requires a detailed description of reionization and evolution of the IGM, which is strongly model dependent. The conclusions drawn here, however, are entirely independent of these details. The colored circles in \Fig{fig:tau} denote the point of production. As is clear, the change in $\tau_{BR}$ at the point of production over a narrow range of $z$ is always large, regardless of the dark photon mass, and consequently we always expect resonantly produced photons to be absorbed instantaneously. Notice that if dark photons are relativistic at conversion, they do not necessarily suffer such large optical depths. Such dark photons cannot themselves constitute dark matter, but, as shown in~\cite{Pospelov:2018kdh}, they may nonetheless have cosmological consequences, such as explaining the anomalously large absorption dip observed in the 21cm spectrum by the EDGES collaboration~\cite{Bowman:2018yin}. 

\begin{figure}[]
\includegraphics[width=0.5\textwidth]{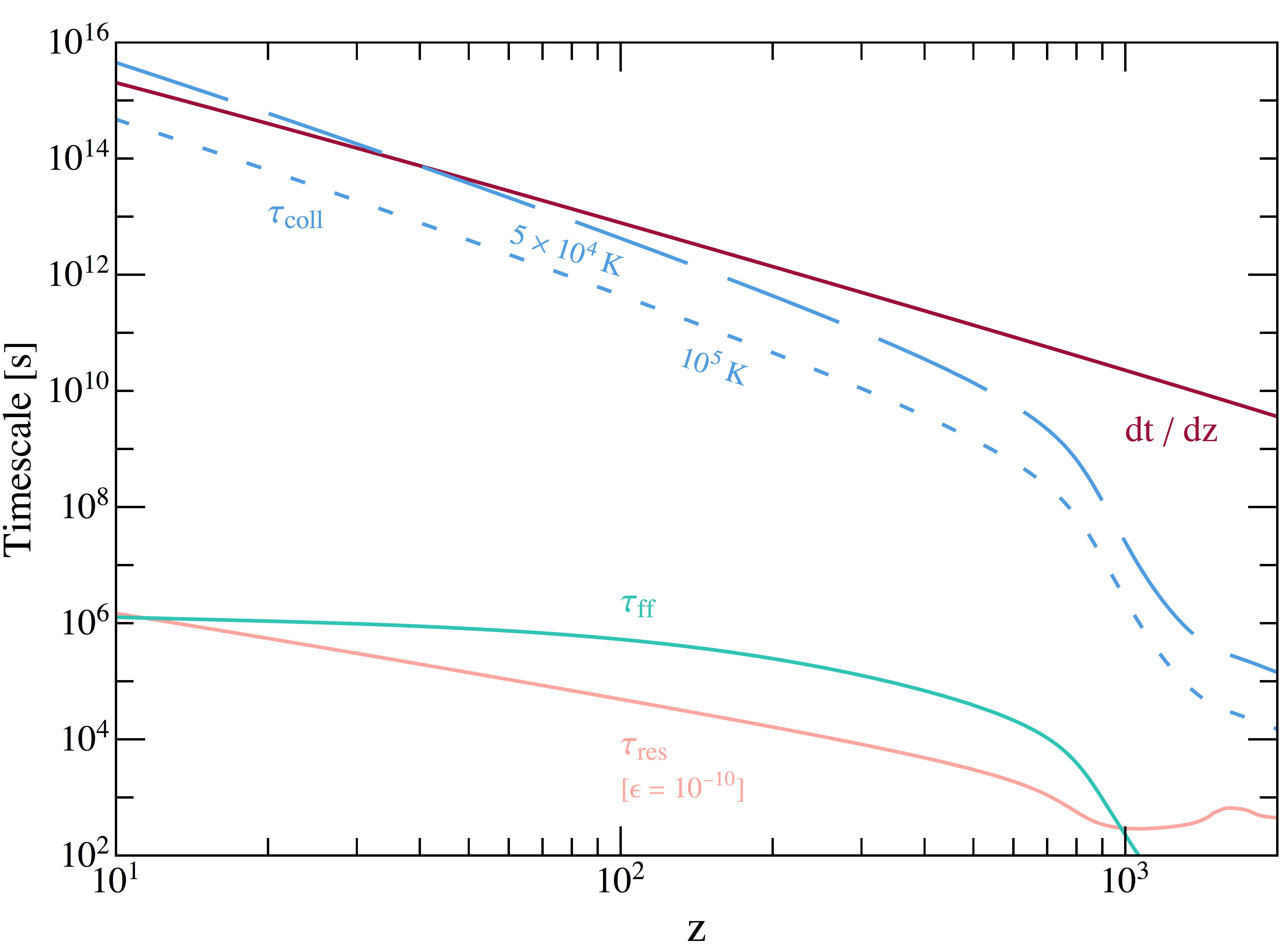}
\caption{Comparison of various timescales relevant for resonant conversion $\tau_{\rm res}$ (assuming $\epsilon = 10^{-10}$), free-free absorption $\tau_{\rm ff}$ (assuming an IGM temperature consistent with $\Lambda$CDM), and collisional ionization of hydrogen $\tau_{\rm coll}$ (assuming an IGM temperature of $5\times10^4$ or $10^5$ K, shown in. long and short dashed lines respectively). We show $dt/dz = 1 / (H(z) \times (1+z))$ for comparison. }
\label{fig:timescales}
\end{figure}

We now turn our attention to understanding the timescales relevant for the injection and absorption of energy, as well as the subsequent ionization. In order to ensure that back reaction is not capable of altering the resonance production of photons, one should verify that
\begin{equation} \label{fast-res}
	\tau_{\rm res} \ll \tau_{\rm ff} + \tau_{\rm coll} \, ,
\end{equation}
where $\tau_{\rm res}$, $\tau_{\rm ff}$, and $\tau_{\rm coll}$ are the characteristic time scales over which the resonance, free-free absorption, and electron-ion collisions take place.  \Eq{fast-res} includes both the free-free and collision times because the resonance condition is sensitive to the value of $\omega_p$, which is a function of $x_e$, but $x_e$ can only change if the plasma is heated and this heating leads to collisional ionization.  The timescale for the resonant transition of dark photons is given by~\cite{Mirizzi:2009iz}
\begin{equation}
\tau_{\rm res} \simeq \left|\frac{d \ln m_\gamma^2(t)}{d t} \right|^{-1}_{t=t_{\rm res}} \times \, \sin(2\epsilon) \,  ,
\end{equation}
while that of free-free absorption is approximately given by
\begin{equation}
\tau_{\rm ff} = \left(\frac{\partial \tau_{BR}(E_\gamma, z)}{\partial\ell}\right)^{-1} \frac{1}{c}
\end{equation}
where we have explicitly included the speed of light dependence for clarity, and $d\ell$ is the differential path of the particle. Should the gas become sufficiently hot, a significant fraction of the gas can undergo collisional ionization. This takes place on timescales 
\begin{equation}
\tau_{\rm coll} = \frac{1}{n_e \times S_{\rm coll}(T)} \, ,
\end{equation}
where $S_{\rm coll}(T)$ is the volumetric collisional ionization rate; for hydrogen, this is approximately given by~\cite{dopita2013astrophysics}
\begin{multline}
S_{\rm coll}(T) \sim 2.5 \times 10^{-10} \left(1 + \frac{T}{78945 \, K} \right) \\ \times \sqrt{T / K} \, e^{-157890 K / T} \, {\rm cm^3 / s} \, .
\end{multline}
In \Fig{fig:timescales} we compare $\tau_{\rm res}$ (with epsilon taken to be $10^{-10}$), $\tau_{\rm ff}$ (computed assuming a temperature given by the mean IGM value in $\Lambda$CDM, noting that larger temperatures correspond to larger timescales), and $\tau_{\rm coll}$ for $T = 5\times10^4$ and $10^5$ K, being characteristic temperatures near which hydrogen should become fully ionized. We compare these times scales with $dt/dz = 1 / (H(z) \times (1+z))$ (red).  \Fig{fig:timescales} clearly illustrates that for the relevant redshifts and parameter space of interest, $\tau_{\rm res} \ll \tau_{\rm ff} + \tau_{\rm coll}$ and thus there is no need to be concerned with the possibility of back reaction on the resonant conversion.  Finally, we must justify that any modeling adopted in this work to account for the relative change in the dark matter energy density or the subsequent heating of the gas is justified. Previously, we had adopted a $\tanh$ function of width $\Delta z = 1$ to model the change in $\rho_{cdm}$; below, we will model with injection and subsequent heating with a gaussian of width $\Delta z = 0.5$. These assumptions should be seen as conservative if  the time scale of this modeling $\Delta t \sim  \mathcal{O}(1) \times \Delta z  / H(z) / (1+z) $ is much greater than either $ \tau_{\rm res}$ (in the case of $\Delta \rho_{cdm}$) or $\tau_{\rm res} + \tau_{\rm ff} $ (in the case of energy injection). \Fig{fig:timescales} clearly illustrates that both of these conditions are always satisfied.

In order to assess the extent to which dark photon resonant transitions enhance the optical depth, we modify the equations tracking the temperature of the medium to include a near-instantaneous energy injection from resonant dark photon conversion. More specifically, we modify the coupled differential system of equations as solved with {\tt Recfast++} tracking the evolution of the ionization fraction of hydrogen and helium, and the temperature of the medium. These equations are given by~\cite{Peebles:1968ja,Seager:1999bc} and are reproduced in our \App{coupled-xe-eqs} for convenience. We also justify in the appendix why it is valid to neglect other cooling contributions that are not typically included in conventional cooling codes, such as free-free cooling, recombination cooling, collisional ionization cooling, and excitation cooling.

\begin{figure}
\includegraphics[width=.48\textwidth]{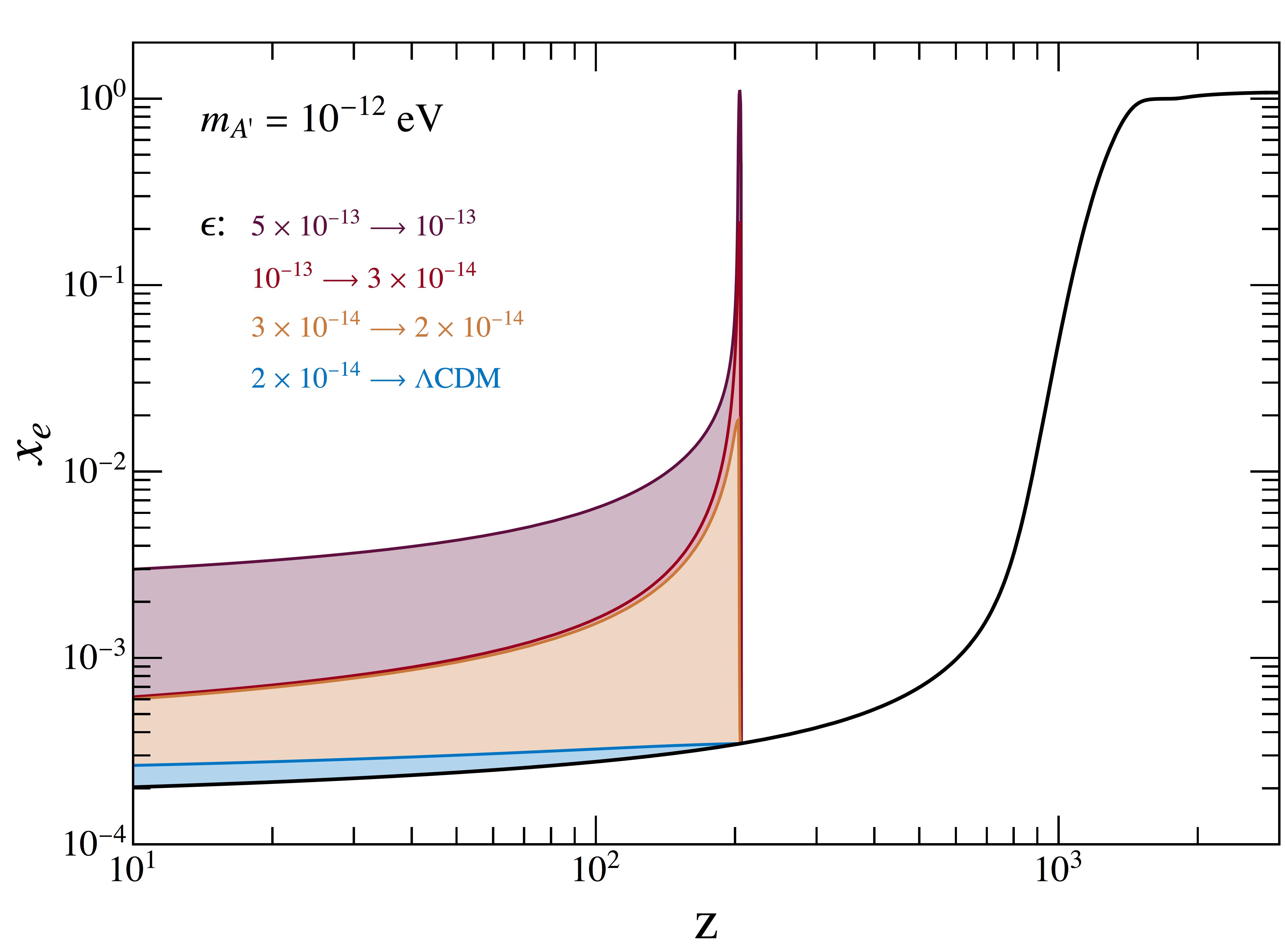}
\caption{\label{fig:xe} Evolution of the free electron fraction for scenarios that account for the resonant conversion of dark photons of mass $m_{A'} = 10^{-12}$ eV and various kinetic mixings. }
\end{figure}

The final term in \Eq{eq:cooling} accounts for the rate at which the plasma is heated via dark photon resonant transitions, where we have implicitly assumed (as justified above) that energy is absorbed on the spot. This term $dE/dz|_{dep}$ is given  by $P_{A'\rightarrow\gamma}^{(\rm res)} \times \rho_{\rm CDM}  \times df(z)/dz $, where we have absorbed the time dependence of the energy injection into the function $f(z)$. Here, we model $f(z)$ as a narrow gaussian centered on the resonance and with a width of $\Delta z = 0.5$. Formally, we include this contribution in the latest version of {\tt Recfast++}~\cite{Seager:1999bc,Chluba:2010ca}, and use this open-source program to determine the evolution of $T_M$. As mentioned before, the energy deposited goes directly into heating the medium; however, once the temperature of the gas is sufficiently high, the gas can become collisionally ionized (this is directly accounted for in \Eqs{eq:newstandard_xe}{eq:HeI_xe}). The evolution of the free-electron fraction must be solved simultaneously with \Eq{eq:cooling}, since these equations are coupled. We illustrate the evolution of the free electron fraction as a function of redshift for a dark photon with mass $m_{A'} = 10^{-12}$ eV  and various mixings in \Fig{fig:xe}. It is clear that the effect of the resonance can be substantial, and is able to significantly increase the integrated optical depth.

We perform a first estimate of this effect by jointly solving for $x_e(z)$ and $T(z)$, as described above, and computing the optical depth by~\cite{Kolb:1990vq,mo2010galaxy,Villanueva-Domingo:2017ahx}
\begin{equation}
\tau = \int \, dz \, \frac{dt}{dz} \, \sigma_T \, n_H^0 (1+z)^3 \left( x_e(z) - x_e^0(z)\right) \, ,
\label{eq:tau}
\end{equation}
where $\sigma_T$ is the Thompson scattering cross section, $n_H^0$ is the number density of hydrogen today, $x_e$ is the free electron fraction as computed here, and $x_e^0$ is the free electron fraction left over after recombination. We include in $x_e$ the effect of late time reionization by astrophysical sources using the $\tanh$ reionization model (see \eg \cite{Villanueva-Domingo:2017ahx}) with a width of 0.5 and a central value of $z = 7$, near the minimum allowed given late time observations of reionization (chosen so as to be maximally conservative). An additional $\tanh$ function is included at $z=3.5$ to account for the second ionization of helium. 

\begin{figure}
\includegraphics[width=.48\textwidth]{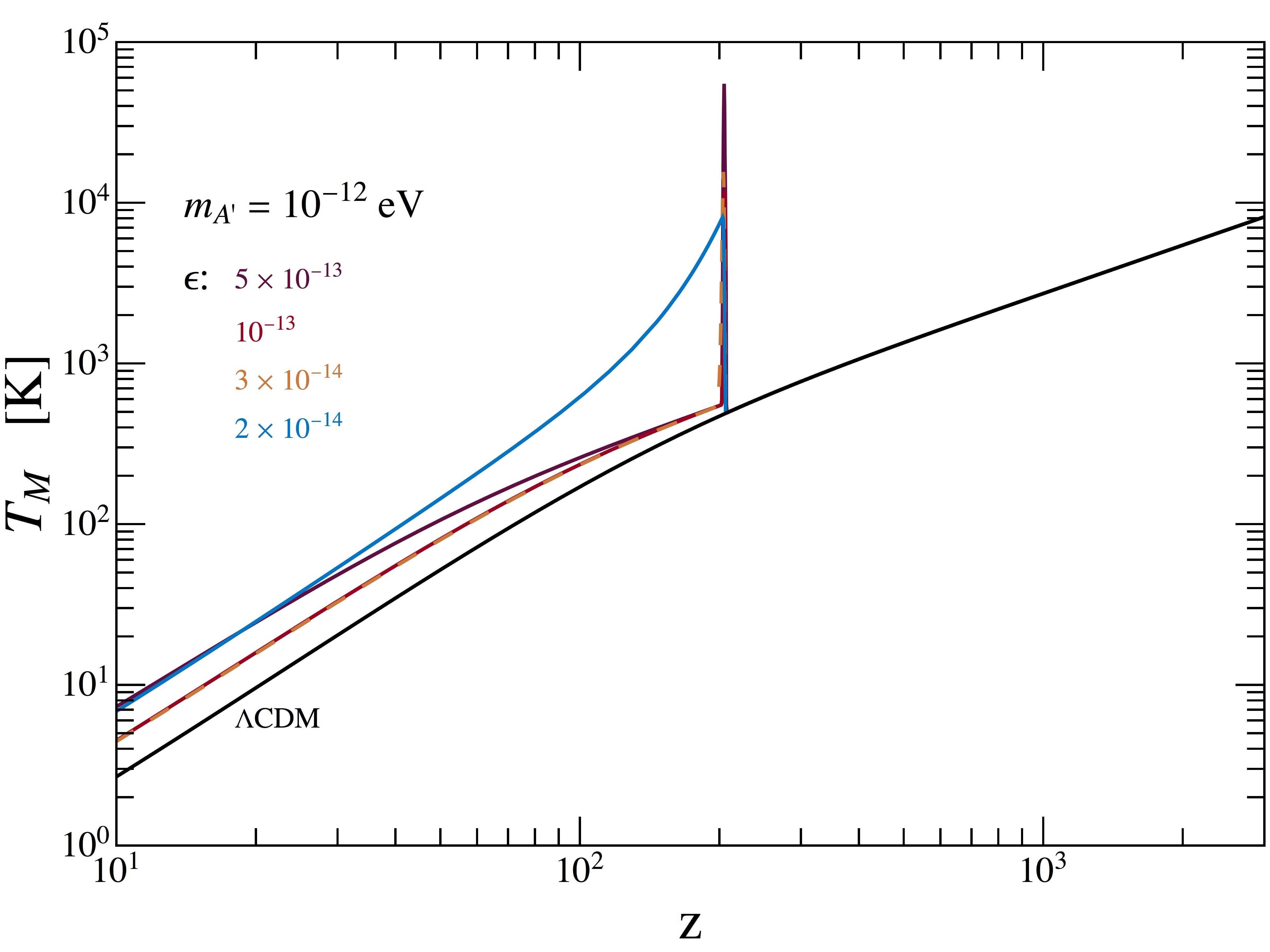}
\caption{\label{fig:tm_evol} Evolution of the matter temperature for the resonant conversion of dark photons of mass $m_{A'} = 10^{-12}$ eV and various kinetic mixings. }
\end{figure}

In order to assess the robustness of this estimate, we  modify {\tt class} to include the effect of heating (in addition to that of dark matter depletion, since these effects must occur simultaneously to be self-consistent). Once again using {\tt montepython}, we perform an MCMC with the {\tt Planck-2018 TTTEEE+low$\ell$TT+lowE+lensing} likelihood~\cite{Aghanim:2019ame}. For models with sufficiently large or late time energy injection, computing the background thermodynamics requires increasing the redshift sampling in {\tt class}. In order to avoid issues with computation speed, we limit our priors on $\log_{10} m_{A'}$ and $\log_{10}\epsilon$ to be between $[-9, -13]$ and $[-12.5, -15]$, respectively. We show the $2\sigma$ bound (labeled `Dark Ages') derived from this analysis (solid) when applicable, and extend to lower masses using the $2\sigma$ bound obtained using only the {\tt Planck} posterior on $\tau$ (dotted), computed using \Eq{eq:tau},  in \Fig{fig:DM} (pink).  These are among the most stringent constraints for dark photons with masses $10^{-14} \lesssim m_{A'} \lesssim 10^{-10}$ eV, losing sensitivity at lower masses  as the effect is masked by astrophysical reionization, and at higher masses by recombination. Notice that while the extension of the contour below the $\oL \omega_p$(today) is perhaps counterintuitive, it is nevertheless correct -- the process of reionization increases the plasma frequency such that $\overline{\omega}_p$(today) is slightly above the pre-reionization value. The primary effect of dark photon is to increase the free electron fraction, thereby increasing the integrated optical depth. In the context of the CMB power spectrum, this appears both as a suppression of the acoustic peaks, and as an enhancement of the low$-\ell$ multipoles in the polarization spectrum. Since both the $TT$ and $EE$ power spectrum constrain the optical depth to comparable levels, we expect both to contribute significantly to the constraining power of the data. For completeness, we also show in \Fig{fig:tm_evol} the evolution of the gas temperature for a $m_{A^\prime} = 10^{-12}$ eV dark photon with various mixing angles. Interestingly, one can see that the asymptotic temperature for small mixings can be comparable to that for large mixings, simply due to the fact that Compton cooling, which cools the gas more rapidly, can dominate over adiabatic cooling for an ionized medium (see Appendix).

Similar to our analyses above, we predict that the heating induced via non-resonant inverse bremsstrahlung may also yield a strong constraint. However, computing this contribution is more complicated than in the case of resonant conversion due to the fact that the frequency of electron-ion collisions will induce a feedback effect: i.e., increasing temperature decreases the rate of energy injection due to the $T_e$ dependence in $\nu$. We estimate that this bound may be a factor of a few stronger than the non-resonant bound derived from helium reionization at masses $m_{A'} \lesssim 10^{-14} \ev$, discussed in the next section, but we leave a rigorous treatment of the implications of non-resonant energy injection in the cosmic dark ages to future work.

\section{Helium II Reionization}
Finally, we address the possibility that dark photon conversion takes place at relatively late times, after baryonic structures have collapsed and UV and X-ray emission from stars and supernovae play an important role in the life of baryons. In particular, we focus on
the epoch in which helium is reionized. Dark photon conversion at this time could lead to an abnormal heating of the IGM. Measurements of the Ly-$\alpha$ forest have been used to infer the temperature evolution of the IGM across the range of redshifts $2 \lesssim z \lesssim 6$. Convincing evidence of a non-monotonic heating of the plasma of the IGM around $z\sim 3.5$ \cite{Schaye:1999vr, Becker:2010cu, Sanderbeck:2015bba} has been interpreted as evidence of the reionization of HeII. Although the magnitude of this feature varies at the $\sim \cO(50\%)$ level in recent analyses \cite{Sanderbeck:2015bba, Hiss:2017qyw, Walther:2018pnn}, a consensus seems strong that the IGM was heated by no more than $\Delta T \lesssim 10^4 \kel \simeq 0.8 \ev$. Since the majority of this heating is surmised to come from the partial ionization of helium atoms, bounds on anomalous heating of the IGM of size $\sim 0.5\ev$ per baryon in the range $2 \leq z \leq 5$ were presented in \cite{Sanderbeck:2015bba}.  Anomalous heating of the IGM on a comparable level can be constrained for redshifts extending to the end of hydrogen reionization, occurring near $z \sim 6$~\cite{bolton2012improved,garzilli2017cutoff}.

In this work, we will impose a conservative limit
\beq \label{HeII}
\Delta \rho_{A'}^{2 \leq z \leq 6} \leq 1\ev \times n_b,
\eeq
where $n_b$ is the total number density of baryons. Only a small fraction of baryons at these redshifts are contained in collapsed objects, so we approximate $n_b$ by the cosmic average~\cite{McQuinn:2015icp}. We consider both resonant and non-resonant absorption of dark photons, as in \Eqs{LZp}{Q-dp}, corresponding to conversion to an on-shell photon or off-shell inverse-bremsstrahlung, respectively. For all dark photon masses $m_{A'} \lesssim 10^{-14} \ev$, this turns out to be the strongest constraint on the dark photon parameter space. Thus, dark photon dark matter that could potentially be heating collapsed structures such as the Milky Way (as suggested by \cite{Dubovsky:2015cca}) or its satellites (as suggested by \cite{Wadekar:2019xnf}) would in fact also have unacceptably heated the IGM at redshift $2 \leq z \leq 6$.

For the range of dark photon masses coinciding with the SM photon plasma mass in this redshift range,  $m_{A'} \sim 10^{-13} \ev$, this bound is stronger than previous cosmological limits \cite{Arias:2012az} by 5 orders of magnitude and stronger than bounds on local collapsed objects \cite{Dubovsky:2015cca, Wadekar:2019xnf} by 4 orders of magnitude. We note that these bounds will scale quadratically in $\ep$, so the bound for $\Delta \rho_{A'} \leq 0.5 \ev \times n_b$ is trivially obtained by rescaling our HeII limit by $\sqrt2$.

\section{Conclusions}
In this work we have revisited cosmological constraints on light to ultralight dark photon dark matter. Since the dark photon mixes with the SM photon, this dark matter candidate is subject to plasma effects such as resonant photon-dark photon conversion and Debye screening, making its phenomenology more diverse than conventional cold dark matter candidates. We have derived novel constraints that cover a far broader mass and mixing range than previously appreciated. We show that very simple and robust cosmological bounds arising from the non-resonant evaporation of dark photons constrain masses as low as $\sim 10^{-20}$ eV. 
Very strong bounds can be attained by requiring dark bremsstrahlung processes not significantly heat the IGM at redshifts for which Ly-$\alpha$ forest measurements probe the epoch of helium reionization (\ie $2 \lesssim z \lesssim 6$). We also demonstrate that resonant bounds derived from helium and post-recombination reionization significantly strengthen existing bounds in the range $10^{-14} \lesssim m_{A'} \lesssim 10^{-9}$ eV. Collectively, the bounds derived here robustly exclude large regions of previously unexplored parameter space for light dark photon dark matter.

One point not directly addressed here, but perhaps worth serious consideration, is the role of plasma inhomogeneities in resonant dark photon conversion. Cosmological studies to date have assumed the plasma frequency is well-characterized by a mean electron number density. This naive assumption likely works quite well when $m_{A'} \sim \oL \omega_p$, where $\oL \omega_p$ indicates the cosmologically averaged value at a given redshift; however, electron under-densities that inevitably exist within the plasma should allow for dark photons with $m_{A'} < \oL \omega_p$ to resonantly convert, a process which is strongly suppressed. The necessary existence of such under-densities implies resonance constraints, typically much stronger than their non-resonant counterparts, extend to a much broader mass range. Depending on the abundance and distribution of these under-densities, it may be possible to derive far more stringent constraints in the low mass regime. We leave the prospect of understanding the role of conversions in inhomogeneities to future work.

~\\ \noindent{\it Acknowledgments:} We would like to thank Prateek Agrawal, Jeff Dror, Olga Mena, Sergio Palomares-Ruiz, Matt Reece, and Lorenzo Ubaldi for various discussions and their comments on the manuscript. SJW acknowledges support under Spanish grants FPA2014-57816-P and FPA2017-85985-P of the MINECO and PROMETEO II/2014/050 of the Generalitat Valenciana, and from the European Union's Horizon 2020 research and innovation program under the Marie Sk\l{}odowska-Curie grant agreements No.\ 690575 and 674896. This manuscript has been authored by Fermi Research Alliance, LLC under Contract No. De-AC02-07CH11359 with the United States Department of Energy.

\bibliography{biblio}

\appendix

\begin{figure}
	\includegraphics[width=0.49\textwidth]{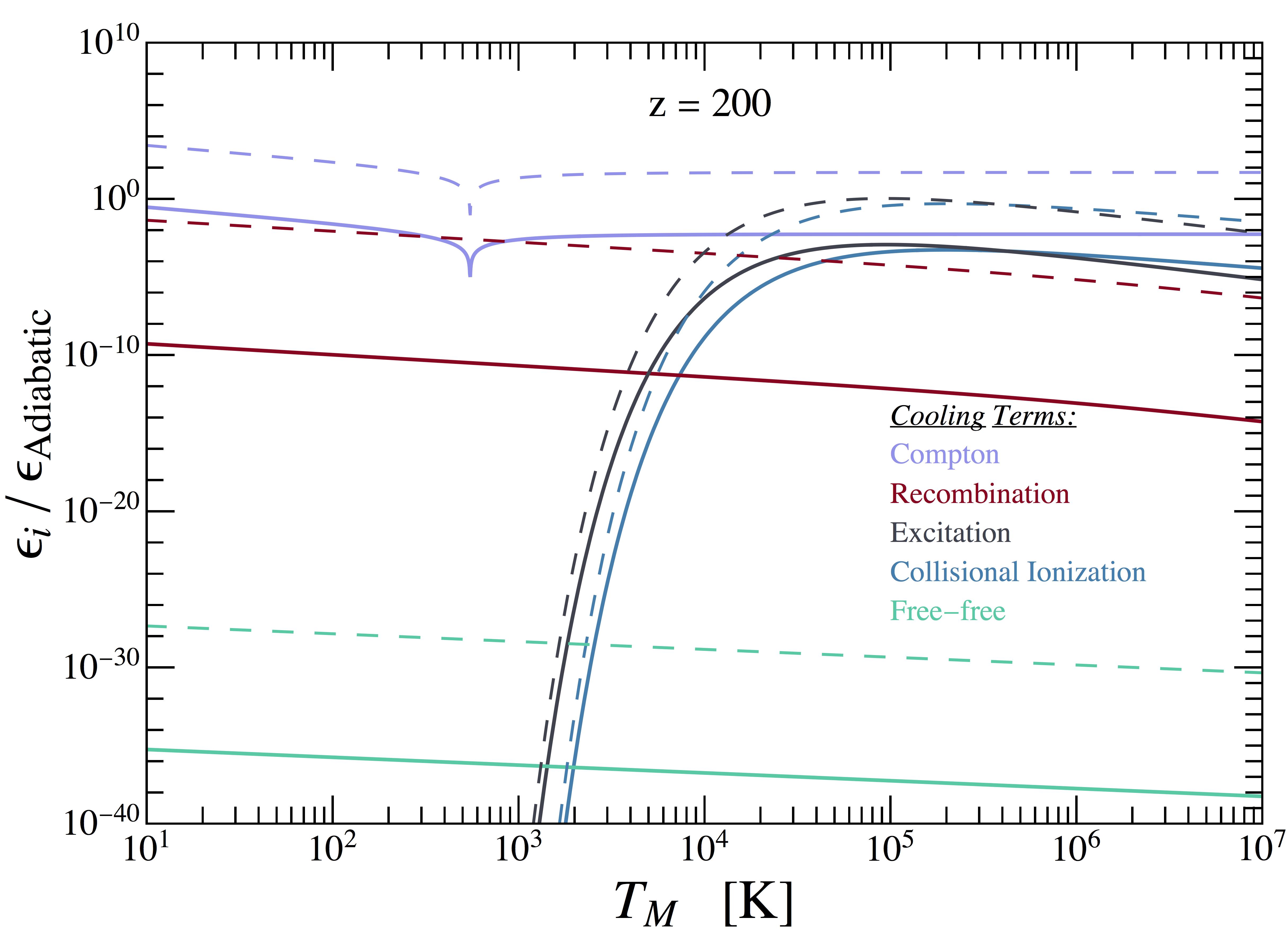}
	\caption{\label{fig:cooling} The ratio of cooling rate $\epsilon_i$ to the adiabatic cooling rate as a function of temperature for the medium at $z=200$, assuming either $x_e = 10^{-4}$ (solid) or $x_e = 0.9$ (dashed).  Processes shown include Compton cooling, collisional ionization cooling, recombination cooling, excitation cooling, and free-free cooling. }
\end{figure}

\section{CMB Spectral Distortions}
Here, we provide a brief description of the origin of spectral distortions of the CMB due to energy injected to or extracted from the SM plasma. The interested reader can find a more extensive discussion in~\cite{Chluba:2011hw}.

Around $z \sim 10^6$, photon production from DC and bremsstrahlung become inefficient at producing high energy photons, although at lower frequencies equilibrium can still be maintained. Compton scattering, however, maintains kinetic equilibrium with the SM plasma; this implies a blackbody spectrum cannot be established. The partial efficiency of thermalization processes are such that the photon distribution can be well-described by a Bose-Einstein distribution with a frequency-dependent chemical potential. For this reason, spectral distortions of this sort are known as $\mu$-type.  At lower redshifts, namely $10^3 \lesssim z \lesssim 10^4$, Compton scattering loses efficiency, implying kinetic equilibrium can no longer be maintained. That is, photons injected from inverse bremsstrahlung tend to stay, at least approximately, locally distributed near the frequencies at which they are injected. This results in a lower (higher) temperature decrement at lower (higher) frequencies, and produces what are known as $y$-type distortions. 

In the epoch between $10^4 \lesssim z \lesssim 10^5$, there exists a complex interplay of processes such that the distortions are not purely $\mu$-type nor $y$-type, but rather a complex admixture. For example, $i-$type distortions, which are distinct from both $\mu$- and $y$-type~\cite{Khatri:2012tw,Khatri:2013dha}, uniquely appear during this epoch. Determining the implications of energy injection during this period on the spectrum typically require a complex numerical study; however, since our formalism neglects $i$-type distortions, we caution the reader that the constraints derived result in a somewhat conservative estimation of the sensitivity.

\begin{widetext}
\section{Equations for the Evolution of the Temperature and the Ionization Fractions}
\label{coupled-xe-eqs}

The following equations solve for the proton fraction $x_{\rm p} = n_{\rm p}/n_{\rm H}$, the fraction of singly ionized helium $x_{\rm HeII} = n_{\rm HeII}/n_{\rm H}$, and the temperature of matter $T_{\rm M}$:
\begin{eqnarray}
\label{eq:newstandard_xe}
{dx_{\rm p}\over dz} &= \left(x_{\rm e}x_{\rm p} n_{\rm H} \alpha_{\rm H}
 - \beta_{\rm H} (1-x_{\rm p})
   {\rm e}^{-h\nu_{H2s}/kT_{\rm M}} - x_e n_H (1-x_p)S_{\rm coll} \right) \\ & \times \frac{\left(1 + K_{\rm H} \Lambda_{\rm H} n_{\rm H}(1-x_{\rm p})\right)}{H(z)(1+z)\left(1+K_{\rm H} (\Lambda_{\rm H} + \beta_{\rm H})
     n_{\rm H} (1-x_{\rm p}) \right)} \nonumber
\end{eqnarray}

\begin{eqnarray}
\label{eq:HeI_xe}
{dx_{\rm He II}\over dz} &=
   \left(x_{\rm He II}x_{\rm e} n_{\rm H} \alpha_{\rm HeI}
   - \beta_{\rm HeI} (f_{\rm He}-x_{\rm He II})
   {\rm e}^{-h\nu_{HeI2^1s}/kT_{\rm M}}\right)  \\  & \times {\left(1 + K_{\rm HeI} \Lambda_{\rm He} n_{\rm H}
  (f_{\rm He}-x_{\rm He II}){\rm e}^{-h\nu_{ps}/kT_{\rm M}})\right)
  \over H(z)(1+z)\left(1+K_{\rm HeI}
  (\Lambda_{\rm He} + \beta_{\rm HeI}) n_{\rm H} (f_{\rm He}-x_{\rm He II})
  {\rm e}^{-h\nu_{ps}/kT_{\rm M}}\right)}, \nonumber
\end{eqnarray}
\begin{equation}
\alpha_{\rm H} = F 10^{-19}\frac{at^{b}}{1 + ct^{d}} \,  \mathrm{m^{3}s^{-1}},
\end{equation}

\begin{equation}
\label{eq:hecoefficient}
\alpha_{\rm HeI} =q\left[\sqrt{T_{\rm M}\over T_2}\left(1+\sqrt{T_{\rm
 M}\over T_2}\right)^{1-p}
 \left(1+\sqrt{T_{\rm M}\over T_1}\right)^{1+p}\right]^{-1}\!
   \mathrm{m^{3}s^{-1}},
\end{equation}

\begin{eqnarray}
\label{eq:cooling}
 \frac{dT_{\rm M}}{dz} &= \frac{8\sigma_{\rm T}a_{\rm R}
   T_{\rm R}^4}{3H(z)(1+z)m_{\rm e}c}\,
  \frac{x_{\rm e}}{1+f_{\rm He}+x_{\rm e}}\,(T_{\rm M} - T_{\rm R}) 
  & + \frac{2T_{\rm M}}{(1+z)} + \frac{2}{3 k}\frac{1}{n_H (1+ f_{He} + x_e)} \left[ \frac{dE}{dz}\Big|_{dep} + \sum_i \epsilon_{i} \right]. 
\end{eqnarray}
\end{widetext}
The electron fraction is then given by $x_{\rm e} = n_{\rm e}/n_{\rm H} = x_{\rm p} + x_{\rm HeII}$, with $n$ being the number density. In the equations above, we have maintained the explicit dependencies on the fundamental constants, such as the speed of light $c$, Boltzmann's constant $k$, Planck's constant $h$, the Thompson scattering cross section $\sigma_T$, the electron mass $m_e$, and the radiation constant $a_R$. The atomic data includes the H Ly$\alpha$ rest wavelength $\lambda_{H2p}=121.5682$nm, the H $2s-1s$ frequency $\nu_{H2s}=c / \lambda_{H2p}$, the He I $2^1p-1^1s$ wavelength $\lambda_{HeI2^1p}=58.4334$nm, the He I $2^1p-1^1s$ frequency $\nu_{HeI2s} = c / 60.1404$nm, with the difference between the two aforementioned being defined as $\nu_{HeI2^1p2^1s} = \nu_{HeI2^1p} - \nu_{HeI2^1s} \equiv \nu_{ps}$, the H $2s-1s$ two photon rate $\Lambda_H = 8.22458 \, s^{-1}$and the He I $2s-1s$ two photon rate $\Lambda_{He} = 51.3 \, s^{-1}$. The $\alpha_H$ case B recombination coefficient for hydrogen contains coefficients $a=4.309$, $b=-0.6166$, $c=0.6703$,  and $d=0.5300$, $t\equiv T_M/10^4$K, and $F=1.14$~\cite{hummer1994total}. The case B recombination coefficient for helium has parameters given by $q=10^{-16.744}$, $p=0.711$, $T_1 = 10^{5.114}$K, and $T_2 = 3$K~\cite{hummer1998recombination}. The $\beta$ factors are the recombinations coefficients, given by $\beta = \alpha (2\pi m_e k T_M / h^2)^{3/2} exp(-h\nu_{2s} / k T_M)$. The cosmological redshifting of H Ly$\alpha$ photons is given by $K_H \equiv \lambda_{H_{2p}}^3 / (8\pi H(z))$, and that of He I $2^1p-1^1s$ is given by $K_{HeI} \equiv \lambda_{HeI_{2^1p}}^3 / (8\pi H(z))$.

The effect of collisional ionization of the ground state of neutral hydrogen has been explicitly included in \Eq{eq:newstandard_xe} following~\cite{Matsuda:1971ry}. We neglect the contribution from the collision excitation and ionization of the excited state for simplicity, as well as the collisional ionization of helium, however these effects are only expected to enhance the asymptotic free electron fraction, and thus the derived constraints.    

The final contributions $\epsilon_i$ in \Eq{eq:cooling} account for all of the possible heating and cooling and processes. In principle, one must be concerned here that after the gas becomes heated, and before ionization, new cooling processes could become active and significantly alter the thermal and ionization properties of the gas. In \Fig{fig:cooling} we show the relative rates of various cooling processes relative to the rate of adiabatic cooling, as a function of the temperature of the medium at $z=200$. For each of the processes shown, we adopt the rates as shown in~\cite{Matsuda:1971ry, Fukugita:1993ds, Thomas:2007dc}. The solid and dashed lines in \Fig{fig:cooling}  depict the rates assuming $x_e = 10^{-4}$ and $x_e = 0.9$, respectively. This plot clearly illustrates that adiabatic and Compton cooling are sufficient to capture the temperature evolution of the gas. It is possible that when generalizing the formalism to include the effects of inhomogeneities that this statement will no longer be valid, and one must be careful in treating high density objects at high temperature.

\end{document}